\documentclass[aps,prl,twocolumn,amsmath,amssymb,showpacs,nofootinbib]{revtex4}
\usepackage{epsfig,float}
\usepackage{graphicx}
\usepackage{dcolumn}
\usepackage{morefloats}
\usepackage{color}
\usepackage{slashed}
\usepackage{bbm}
\usepackage{epstopdf}

\begin{document}

\title{Could the near--threshold $XYZ$ states be simply kinematic effects?}

\author{Feng-Kun Guo$^1$\footnote{{\it Email address:} fkguo@hiskp.uni-bonn.de}, Christoph Hanhart$^2$\footnote{{\it Email address:}
        c.hanhart@fz-juelich.de}, Qian Wang$^2$\footnote{{\it Email address:} q.wang@fz-juelich.de},
 Qiang Zhao$^3$\footnote{{\it Email address:}
        zhaoq@ihep.ac.cn} }

\affiliation{$^1$ Helmholtz-Institut f\"ur Strahlen- und Kernphysik and
Bethe Center for Theoretical Physics, Universit\"{a}t Bonn, D-53115 Bonn, Germany\\
       $^2$ Institut f\"{u}r Kernphysik and  Institute for Advanced Simulation,
          Forschungszentrum J\"{u}lich, D--52425 J\"{u}lich, Germany\\
       $^3$ Institute of High Energy Physics and Theoretical Physics Center for Science Facilities,
        Chinese Academy of Sciences, Beijing 100049, China}

\begin{abstract}

We demonstrate that the spectacular structures discovered recently  in various
experiments and named as $X$, $Y$ and $Z$ states cannot be purely kinematic
effects. Their existence necessarily calls for nearby poles in the $S$--matrix
and they therefore qualify as states. We propose a way to distinguishing kinematic cusp effects from genuine $S$--matrix poles:
 the kinematic threshold cusp cannot produce a narrow peak in the invariant mass distribution in the elastic channel in contrast with a genuine
 $S$--matrix pole.

\end{abstract}

\date{\today}


\maketitle

In recent years various narrow  (widths from well below 100 MeV down to values even below 1 MeV)
peaks were discovered both in the charmonium
as well as in the bottomonium mass range that do not fit into the so far very
successful quark model. For instance, the most prominent ones include
$X(3872)$~\cite{Choi:2003ue},
$Z_c(3900)$~\cite{Ablikim:2013mio,Liu:2013dau,Xiao:2013iha},
$Z_c(4020)$~\cite{Ablikim:2013xfr,Ablikim:2013wzq,Ablikim:2013emm,Ablikim:2014dxl},
$Z_b(10610)$ and $Z_b(10650)$~\cite{Belle:2011aa}, which are located close to
$D\bar D^*$, $D\bar D^*$, $D^*\bar D^*$, $B\bar B^*$ and  $B^*\bar B^*$
thresholds in relative $S$--waves, respectively. Apart from  other
interpretations, such as
hadro-quarkonia~\cite{Voloshin:2007dx,Dubynskiy:2008mq},
hybrids~\cite{Zhu:2005hp,Kou:2005gt,Close:2005iz}, and
tetraquarks~\cite{Maiani:2014aja,Faccini:2013lda} (for recent reviews we refer
to Refs.~\cite{Brambilla:2010cs,italians}),  due to their proximity to the
thresholds these five states were proposed to be of a molecular
nature~\cite{Tornqvist:2004qy,Fleming:2007rp,Thomas:2008ja,Ding:2008gr,Lee:2009hy,Dong:2009yp,Stapleton:2009ey,
Gamermann:2009fv,Mehen:2011ds,Bondar:2011ev,
Nieves:2011vw,Nieves:2012tt,
Wang:2013cya,Wang:2013kra,Guo:2013sya,Guo:2013zbw,
Mehen:2013mva,He:2013nwa,Liu:2014eka}.
As an alternative explanation various groups conclude from the mentioned
proximity of the states to the thresholds that the structures are simply
kinematical
effects~\cite{Bugg:2004rk,Bugg:2011jr,Chen:2011pv,Chen:2011xk,Chen:2011pu,Chen:2013coa,Chen:2013wca,Swanson:2014tra}
that necessarily occur near every $S$-wave threshold.  Especially, it has been
claimed that the structures are not related to a pole in the $S$--matrix and
therefore should not be interpreted as states.

In this letter we show that the latter statement is  based on calculations
performed within an inconsistent formalism.
In particular, we demonstrate that, while there is always a cusp at the opening
of an $S$--wave threshold, in order to produce peaks as pronounced and narrow as
observed in experiment non-perturbative interactions amongst the heavy mesons
are necessary, and as a consequence, there is to be a near-by pole.
Or, formulated the other way around: if one assumes the two--particle
interactions to be perturbative, as it is implicitly done in
Refs.~\cite{Bugg:2004rk,Bugg:2011jr,Chen:2011pv,Chen:2011xk,Chen:2011pu,Chen:2013coa,Chen:2013wca,Swanson:2014tra},
the cusp should not appear as a prominent narrow peak.
This statement is probably best illustrated by the famous $K^\pm\to
\pi^\pm\pi^0\pi^0$ data~\cite{Batley:2000zz}:
the cusp that appears in the $\pi^0\pi^0$ invariant mass distribution at the
$\pi^+\pi^-$ threshold is a very moderate kink, since the $\pi\pi$ interactions
are sufficiently weak to allow for a perturbative treatment (for a comprehensive
theoretical framework and related references we refer to
Ref.~\cite{Gasser:2011ju}).

\begin{figure}[t] \vspace{0.cm}
\begin{center}
\hspace{2cm}
\includegraphics[scale=0.6]{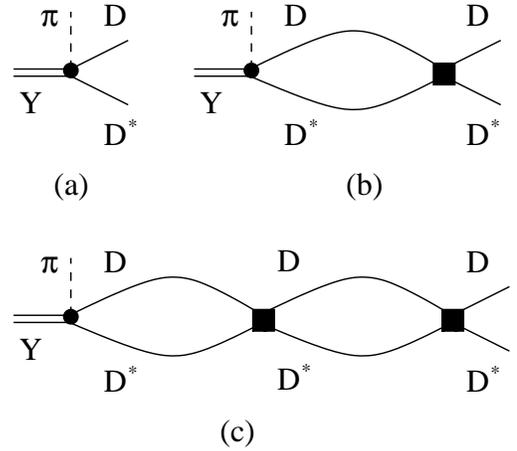}
\caption{The tree-level, one-loop and two-loop Feynman diagrams for $Y(4260)\to \pi D \bar D^*$.}
 \label{diagrams1}
\end{center}
\end{figure}

To be concrete, in this paper we demonstrate our argument on the example
of an analysis of the existing data on the $Z_c(3900)$, but it should be clear
that the reasoning as such is general and applies to all  structures observed
very near
$S$--wave thresholds such as those above-mentioned $XYZ$ states.
To illustrate our point, we here do not aim for field theoretical rigor
but use a very simple separable interaction for all vertices
accompanied by loops
regularized with a Gaussian regulator. This regulator will at the
same time control the drop-off of the amplitudes as will be discussed below.
Accordingly, we write for the Lagrangian that produces the tree--level vertices
(here and in what
follows we generically write $D\bar D^*$ for the proper
linear combination of $D\bar D^*$ and $\bar DD^*$)
\begin{eqnarray}\nonumber
{\cal L}_I&=&g_Y  \pi (D\bar D^*_\mu)^\dagger Y^\mu + \frac{C}{2} (D\bar D^*)^\dagger(D\bar D^*) \\
& & \quad   + \, g_{\psi Y}\psi^{\mu\dagger}\pi\pi Y_{\mu }  + g_\psi
\psi^{\mu \dagger} \pi D\bar D^*_\mu + ... \ ,
\end{eqnarray}
where $Y$, $D$, $D^*$, $\pi$ and $\psi$ denote the fields for the
$Y(4260)$, $D$, $D^*$, $\pi$
and $J/\psi$, respectively.  The dots indicate terms not needed for this study
like the one where the $Y$-field is created.
All fields but the pion field are nonrelativistic and accordingly the
couplings $g_Y$ and $g_\psi$ have
dimension GeV$^{-3/2}$, $g_{\psi Y}$ has dimension GeV$^{-1}$, while $C$ has
dimension GeV$^{-2}$.
The loops are regularized with the cutoff function $f_\Lambda(\vec p\, ^2)$,
which for convenience we choose as
\begin{equation}
f_\Lambda(\vec p\, ^2) = \exp\left(-2\vec p\, ^2/\Lambda^2\right) \ ,
\label{eq:ff}
\end{equation}
where here and below $\vec p$ denotes the three-momentum of the $D$-meson in
the center-of-mass frame of the $D\bar D^*$ system.
Therefore the loop function reads
\begin{equation}
G_\Lambda(E)=\int \frac{d^3 q}{(2\pi)^3}\frac{f_\Lambda(\vec q\, ^2)}{E-m_1
-m_2 -\vec q\, ^2/(2\mu)} \ ,
\end{equation}
where $m_{1,2}$ denote the masses of the charmed mesons, $\mu$ is the reduced
mass and $E$ is the total energy. With the regulator specified in
Eq.~\eqref{eq:ff}, the analytic expression for the loop function for $E\geq
m_1+m_2$ is given by
\begin{eqnarray}
G_\Lambda(E) = -\frac{ \mu\Lambda }{(2\pi)^{3/2}} + \frac{\mu k}{2\pi}
e^{-2k^2/\Lambda^2} \left[ \text{erfi}\left(\frac{\sqrt{2}k}{\Lambda}\right) - i
\right],
\label{eq:gexplicit}
\end{eqnarray}
where $k = \sqrt{2\mu (E-m_1-m_2)}$, and
\begin{equation}
\text{erfi}(z) = \frac2{\sqrt{\pi}}
\int_0^z e^{t^2}dt
\end{equation}
is the imaginary error function.

\begin{figure}[t] \vspace{0.cm}
\begin{center}
\hspace{2cm}
\includegraphics[scale=0.6]{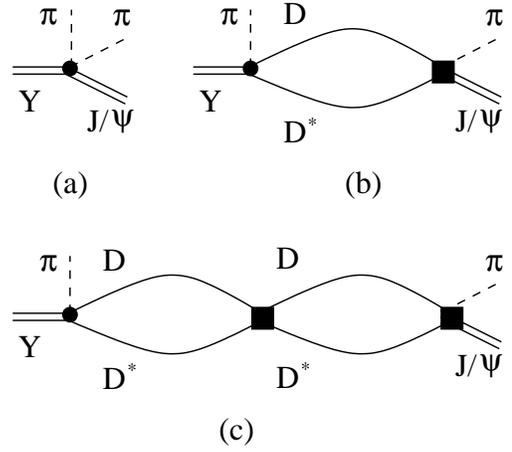}
\caption{The tree-level, one-loop and two-loop Feynman diagrams for $Y(4260)\to \pi\pi J/\psi$.}
 \label{diagrams2}
\end{center}
\end{figure}

With the ingredients of the model fixed it is straightforward to derive the
explicit expressions for the transition matrix elements. Within this model
the $Y(4260)\to \pi D\bar D^*$ amplitude reads to one-loop order
($cf$. the diagrams of Fig.~\ref{diagrams1} (a)+(b))
\begin{equation}
g_Y \left[1 - G_\Lambda(E)C \right] \ .
\label{eq:ddstarpi}
\end{equation}
The analogous result for the $Y(4260)\to \pi \pi J/\psi$ amplitude is
($cf$. the diagrams of Fig.~\ref{diagrams2} (a)+(b))
\begin{equation}
g_{\psi Y} - g_Y G_\Lambda(E) g_\psi \  .
\label{eq:jpsipipi}
\end{equation}

\begin{figure}[t!] \vspace{0.cm}
\begin{center}
\includegraphics[width=\linewidth]{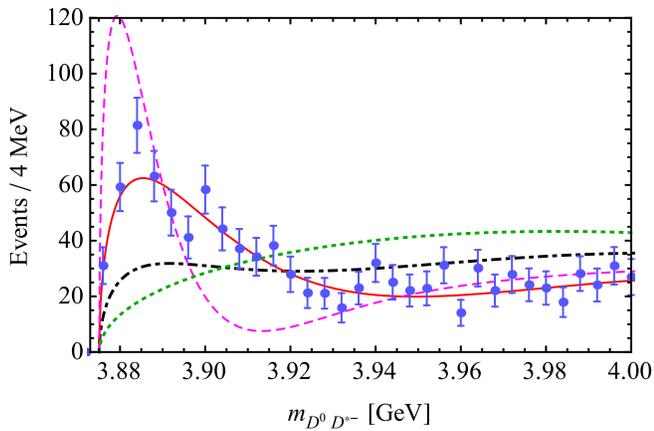}
\caption{Results for the $D\bar D^*$ invariant mass distribution in
$Y(4260)\to \pi D \bar D^*$. The data are from Ref.~\cite{Ablikim:2013xfr} and
the results from the tree level, full one-loop and full two-loop
calculations are shown by the dotted, solid and dashed curves, respectively.
The dot--dashed line shows the one-loop result with the strength
of the rescattering requested to be small to justify a perturbative treatment
as described in the text.}
 \label{DDst}
\end{center}
\end{figure}

We now proceed as follows: We first confirm the claims of Refs.~\cite{Bugg:2011jr,Chen:2013coa,Swanson:2014tra},
namely, that the data available for both $Y(4260)\to \pi D\bar D^*$
as well as $Y(4260)\to \pi\pi J/\psi$
can at least qualitatively be described by a sum of the tree-level and
one--loop diagrams shown in Fig.~\ref{diagrams1} (a)+(b)
and Fig.~\ref{diagrams2} (a)+(b), respectively.
Note that diagram (b) in either Fig.~\ref{diagrams1} or \ref{diagrams2}
explicitly contains the above mentioned cusp. It was this observation that lead the authors
 of Refs.~\cite{Bugg:2011jr,Chen:2013coa,Swanson:2014tra} to interpret the
 near--threshold structures as purely kinematical effect.
To fix the parameters we first fix $g_Y$, $\Lambda$ and $C$ by a fit to
the $D\bar D^*$ spectrum. The fit result is shown by the solid line in
Fig.~\ref{DDst} (the corresponding strength of the tree level diagram
is shown by the dotted line). In particular we find
\begin{eqnarray}
\label{para}
 C= 64.4~\text{GeV}^{-2}, \quad \Lambda= 0.326~\text{GeV}
\end{eqnarray}
and $g_Y=102.6$~GeV$^{-3/2}$ (notice that this parameter is not normalized to
the physical value since we are fitting to the number of events, and a factor of
$\sqrt{8 m_Y m_D m_{D^*}}$ needs to be multiplied to it in order to obtain the
solid curve shown in Fig.~\ref{DDst}) for the best fit.
It is crucial for the reasoning of this letter that the contribution
from the tree--level source term ($cf$. Fig.~\ref{diagrams1}(a)) and the $D\bar D^*$ rescattering
($cf$. Fig.~\ref{diagrams1}(b))
can be disentangled, since the former is fixed by the $D\bar D^*$ spectrum for
values of $m_{D\bar D^*}$   above around
3.94~GeV, while the latter is to explain the structure for values
below this invariant mass, see
Fig.~\ref{DDst}.

\begin{figure}[t!] \vspace{0.cm}
\begin{center}
\includegraphics[width=\linewidth]{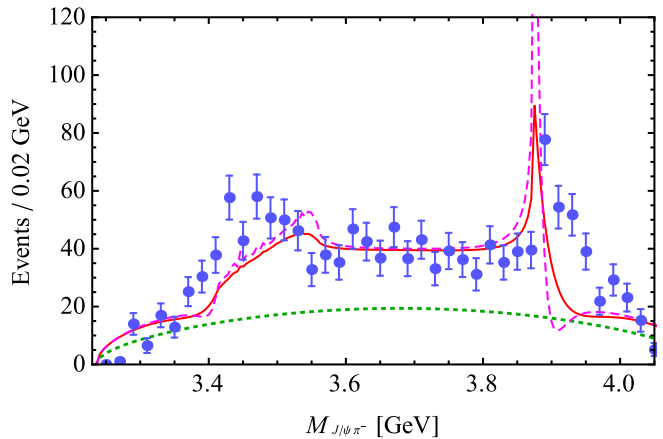}
\caption{Results for the $\pi J/\psi$ invariant mass distribution in
$Y(4260)\to \pi \pi J/\psi$. The data are from Ref.~\cite{Ablikim:2013mio} and
the results from the tree level, full one-loop and full two-loop
calculations are shown by the dotted, solid and dashed curves, respectively,
with the cutoff as well as $g_Y$ from the fit to the $D\bar D^*$ spectrum. }
 \label{pipiJpsi}
\end{center}
\end{figure}

Next we keep $g_Y$ and $\Lambda$ fixed and fit $g_\psi$ and $g_{\psi Y}$ to the
$J/\psi \pi$ spectrum. The best fit gives $g_{\psi Y} = 46.4$~GeV$^{-3/2}$ and
$g_{\psi}=0.44$~GeV$^{-3/2}$ which are also not normalized to the physical
values due to fitting to the event numbers.
The result of this fit is shown  as the solid line in Fig.~\ref{pipiJpsi}.
In this work we only aim at a qualitative description of the data.
It should be mentioned that we can get a perfect fit of the $J/\psi\pi$
spectrum, if we also fit $\Lambda$, but then we have to compromise on the fit
quality for the $D\bar D^*$ channel\footnote{Note that the cut--off function $f_\Lambda(\vec p\, ^2)$ is
needed in phenomenological studies not only to regularize the real parts of the loops,
 but also to tame the size especially of the imaginary parts
that would keep rising otherwise.  In this way $f_\Lambda(\vec p\, ^2)$ controls
the shape of the peaks calculated in the model.}.
Since this fitting procedure leads us to the same conclusions we do not show the corresponding fit results.

As mentioned above, the intrinsic assumption of the approaches outlined
in Refs.~\cite{Bugg:2011jr,Chen:2013coa,Swanson:2014tra}  is that the interactions are perturbative, and
consequently, the amplitude is properly represented by the one loop
result. With the parameters fixed we can now calculate
the amplitudes to two--loop order from
\begin{equation}
g_Y \left[ 1 - G_\Lambda(E)C + (G_\Lambda(E)C)^2 \right] \ ,
\end{equation}
for the $\pi D\bar D^*$  channel ($cf$. Fig.~\ref{diagrams1} (a)+(b)+(c))  and
\begin{equation}
g_{\psi Y} - g_YG_\Lambda(E)g_\psi + g_yG_\Lambda(E)CG_\Lambda(E)g_\psi \
\end{equation}
for the $\pi\pi J/\psi$ channel  ($cf$. Fig.~\ref{diagrams2} (a)+(b)+(c)). The results are shown
 as the dashed lines in Figs.~\ref{DDst} and \ref{pipiJpsi}, respectively.
As one can see, in both cases the two-loop result significantly deviates from the
one-loop result around the peak,  which clearly calls for a resummation of the
series.

In fact, when we sum all loops in the $D\bar D^*$ channel using the parameters
of Eqs.~(\ref{para}), the series produces a bound state pole right below
threshold.~\footnote{{In order to search for a pole below
threshold in the first Riemann sheet, we need to analytically continue the expression of the
loop function given in Eq.~\eqref{eq:gexplicit}. This can be done, e.g., by
replacing $E$ by $E+i\epsilon$, where $\epsilon$ is an infinitesimal positive
number.}} This means the following for the results of
Refs.~\cite{Bugg:2011jr,Chen:2013coa,Swanson:2014tra}: if one wants to fit the available data for the near-threshold $Z_c(3900)$ states within a perturbative approach, the presence of a pronounced near-threshold structure calls for such a large coupling constant  that the use of a perturbative approach is not justified. This demonstrates explicitly that the approach used in
Refs.~\cite{Bugg:2011jr,Chen:2013coa,Swanson:2014tra}  is intrinsically
inconsistent.

This argument also works in the other direction: we may constrain the coupling
$C$ for $D\bar D^*$ elastic scattering to a value where it might still be
justified to treat $D\bar D^*$ scattering perturbative, e.g. one may require in
the full kinematic regime $|G_\Lambda(E)C|\ll 1$. Since $|G_\Lambda(E)|$
is maximal for $E=2M$, we may demand  $|C\, G_\Lambda(m_1+m_2)| = a$ with $a\ll
1$.
For $\Lambda$ as given in Eq.~(\ref{para}) and $a=1/2$ we can
again calculate the amplitude to one--loop order. The resulting $D\bar D^*$
spectrum is shown by the dot--dashed line in Fig.~\ref{DDst}. Clearly,
such a small coupling is not able to produce the pronounced structure
in the data.

In the calculation described above we used a Gaussian form factor to regularize
the loop. We checked that a different regulator leads to qualitatively similar
results. Especially the conclusions stay unchanged. In fact,  any other
form factor which is commonly used drops off more slowly for higher momenta.
As a result an even larger value of $|C\, G_\Lambda(m_1+m_2)|$ will be connected
to a narrow near--threshold structure. From this point of view, the
use of a Gaussian form factor as employed above already leads to the most
conservative estimate of the higher loop effects. We should also
mention that the contact interaction and the regularized loop function always
appear in a product, i.e.
$G_\Lambda(E)C$, so that the momentum dependence introduced in the cut-off
function can be equivalently regarded as momentum dependence in the
interaction.

To distinguish an $S$--matrix pole from a simple cusp
effect it is necessary to fix the strength of
the production vertex and of the meson--meson rescattering separately.
This is possible only for the elastic channel, as can be clearly seen from
comparing Eqs.~\eqref{eq:ddstarpi} and \eqref{eq:jpsipipi}: the term $C
G_\Lambda(E)$ which controls the elastic interaction strength can be fixed from
the peak since it interferes with 1, while the inelastic coupling strength
$g_\psi$ in Eq.~\eqref{eq:jpsipipi} always appears in a product with $g_Y$.
We therefore strongly urge all groups claiming a purely kinematic
origin of some near threshold structure to also calculate the transition
of that structure into the corresponding continuum channel and follow
the steps of this paper to either confirm or disprove their claim.

One may wonder if triangle singularities are capable to circumvent 
the argument presented in this paper. After all they are in principle
able to provide enhancements in observables as demonstrated
in a different context, e.g. in Refs.~\cite{Wu:2011yx,Wu:2012pg} (for
 a recent discussion see Ref.~\cite{Szczepaniak:2015eza}).
 However,
this mechanism is effective only in a very limited
kinematic regime and therefore operative only for
selected transitions. Therefore, the very fact that e.g. the $X(3872)$ is seen, amongst others, in
$B\to KX$ and $Y(4260)\to \gamma X$ is a clear
indication that its existence is not exclusively driven
by a triangle singularity. For the case of the $Z_c(3900)$ the
dependence of the triangle singularity on the external energies
is discussed in Ref.~\cite{Wang:2013hga}\footnote{Note: in this work
as well as in Ref.~\cite{Wang:2013cya} the triangle singularity and an explicit pole were included
simultaneously.}.
Probably even more  important for the line of reasoning
in this paper, for the elastic channel, i.e. when the
incoming and outgoing particles in the final state interaction as part of the
triangle diagram are the same, the triangle singularity will not produce any
peak~\cite{Schmid:1967,Anisovich:1995ab}. Thus, our conclusion which relies 
mainly on
the analysis in the elastic (continuum) channel remains: pronounced, narrow
near-threshold peaks cannot be produced by purely kinematic effects.

Although in this work all calculations are tuned to the production of the
$Z_c(3900)$ seen in $Y(4260)\to \pi Z_c(3900)$ it should be understood that the
arguments are indeed very general:
any consistent treatment of the spectacular  very near-threshold
structures, namely some of those $XYZ$ states, necessarily needs the inclusion
of a nearby pole, which was done, e.g., in
Refs.~\cite{Voloshin:2007dx,Dubynskiy:2008mq,Zhu:2005hp,Kou:2005gt,Close:2005iz,
Maiani:2014aja,Faccini:2013lda,
Tornqvist:2004qy,Fleming:2007rp,Thomas:2008ja,Ding:2008gr,Lee:2009hy,Dong:2009yp,Stapleton:2009ey,
Gamermann:2009fv,Mehen:2011ds,Bondar:2011ev,
Nieves:2011vw,Nieves:2012tt,
Wang:2013cya,Wang:2013kra,Guo:2013sya,Guo:2013zbw,
Mehen:2013mva,He:2013nwa,Liu:2014eka,Danilkin:2011sh}.
 For each individual state
  a detailed high--quality fit to the data is necessary to decide
if this pole is located on the first sheet (bound state) or on the second sheet
(virtual state or resonance).
It also requires additional research to decide on the origin of that pole, which
might, e.g., come from short--ranged four--quark interactions or from
meson--meson interactions. All we can conclude from the results of this paper is
that there has to be a near--threshold pole.

%\vspace{0.2cm}

We are grateful for the inspiring atmosphere at the
Quarkonium Working Group Workshop 2014 where the idea for this
work was born as well as to very useful discussions with Eric Braaten, Estia Eichten,
Tom Mehen, Ulf-G. Mei{\ss}ner and Eric Swanson.
This work is supported, in part, by NSFC and DFG
through funds provided to the Sino-Germen CRC 110 ``Symmetries and
the Emergence of Structure in QCD" (NSFC Grant No. 11261130311),
NSFC (Grant Nos. 11035006 and 11165005), the Chinese Academy of
Sciences (KJCX3-SYW-N2),  and the Ministry of Science and Technology of China
(2015CB856700).


\begin{thebibliography}{99}


%\cite{Choi:2003ue}
\bibitem{Choi:2003ue}
  S.~K.~Choi {\it et al.}  [Belle Collaboration],
  %``Observation of a narrow charmonium - like state in exclusive B+- ---> K+- pi+ pi- J / psi decays,''
  Phys.\ Rev.\ Lett.\  {\bf 91}, 262001 (2003)  [hep-ex/0309032].

%\cite{Ablikim:2013mio}
\bibitem{Ablikim:2013mio}
  M.~Ablikim {\it et al.}  [BESIII Collaboration],
  %``Observation of a charged charmoniumlike structure in e+e- to pi+pi-J/psi at \sqrt{s}=4.26 GeV,''
  Phys.\ Rev.\ Lett.\  {\bf 110}, 252001 (2013)
  [arXiv:1303.5949 [hep-ex]].
  %%CITATION = ARXIV:1303.5949;%%
  %120 citations counted in INSPIRE as of 07 Mar 2014


 %\cite{Liu:2013dau}
\bibitem{Liu:2013dau}
  Z.~Q.~Liu {\it et al.}  [Belle Collaboration],
  %``Study of $e^+ e^- \to \pi^+ \pi^- J/\psi$ and Observation of a Charged Charmonium-like State at Belle,''
  Phys.\ Rev.\ Lett.\  {\bf 110}, 252002 (2013)
%  [arXiv:1304.0121 [hep-ex]].
  %%CITATION = ARXIV:1304.0121;%%
  %97 citations counted in INSPIRE as of 26 Feb 2014


%\cite{Xiao:2013iha}
\bibitem{Xiao:2013iha}
  T.~Xiao, S.~Dobbs, A.~Tomaradze and K.~K.~Seth,
  %``Observation of the Charged Hadron $Z_c^{\pm}(3900)$ and Evidence for the Neutral $Z_c^0(3900)$ in $e^+e^-\to \pi\pi J/\psi$ at $\sqrt{s}=4170$ MeV,''
  Phys.\ Lett.\ B {\bf 727}, 366 (2013)
  [arXiv:1304.3036 [hep-ex]].
  %%CITATION = ARXIV:1304.3036;%%
  %67 citations counted in INSPIRE as of 07 Mar 2014

%\cite{Ablikim:2013xfr}
\bibitem{Ablikim:2013xfr}
  M.~Ablikim {\it et al.}  [BESIII Collaboration],
  %``Observation of a charged (DD*bar)- mass peak in e+e- --> pi+ (DD*bar)- at Ecm=4.26 GeV,''
  Phys.\ Rev.\ Lett.\  {\bf 112}, 022001 (2014)
  [arXiv:1310.1163 [hep-ex]].
  %%CITATION = ARXIV:1310.1163;%%
  %13 citations counted in INSPIRE as of 07 Mar 2014




 %\cite{Ablikim:2013wzq}
\bibitem{Ablikim:2013wzq}
  M.~Ablikim {\it et al.}  [BESIII Collaboration],
  %``Observation of a charged charmoniumlike structure Z_c(4020) and search for the Z_c(3900) in e+e- to pi+pi-h_c,''
  Phys.\ Rev.\ Lett.\  {\bf 111}, 242001 (2013)
  [arXiv:1309.1896 [hep-ex]].
  %%CITATION = ARXIV:1309.1896;%%
  %34 citations counted in INSPIRE as of 07 Mar 2014

\bibitem{Ablikim:2013emm}
  M.~Ablikim {\it et al.}  [BESIII Collaboration],
  %``Observation of a charged charmoniumlike structure in $e^+e^- \to (D^{*} \bar{D}^{*})^{\pm} \pi^\mp$ at $\sqrt{s}=4.26$GeV,''
  arXiv:1308.2760 [hep-ex].
  %%CITATION = ARXIV:1308.2760;%%
  %41 citations counted in INSPIRE as of 07 Mar 2014

%\cite{Ablikim:2014dxl}
\bibitem{Ablikim:2014dxl}
  M.~Ablikim {\it et al.}  [BESIII Collaboration],
  %``Observation of $e^+e^-\to \pi^0\pi^0 h_c$ and a neutral charmoniumlike structure $Z_c(4020)^0$,''
  arXiv:1409.6577 [hep-ex].





%\cite{Belle:2011aa}
\bibitem{Belle:2011aa}
  A.~Bondar {\it et al.}  [Belle Collaboration],
  %``Observation of two charged bottomonium-like resonances in Y(5S) decays,''
  Phys.\ Rev.\ Lett.\  {\bf 108}, 122001 (2012)
  [arXiv:1110.2251 [hep-ex]].
  %%CITATION = ARXIV:1110.2251;%%
  %131 citations counted in INSPIRE as of 07 Mar 2014



\bibitem{Voloshin:2007dx}
  M.~B.~Voloshin,
  %``Charmonium,''
  Prog.\ Part.\ Nucl.\ Phys.\  {\bf 61}, 455 (2008).
%  [arXiv:0711.4556 [hep-ph]].

\bibitem{Dubynskiy:2008mq}
  S.~Dubynskiy and M.~B.~Voloshin,
  %``Hadro-Charmonium,''
  Phys.\ Lett.\ B {\bf 666}, 344 (2008).
%  [arXiv:0803.2224 [hep-ph]].




\bibitem{Zhu:2005hp}
  S.-L.~Zhu,
  %``The Possible interpretations of Y(4260),''
  Phys.\ Lett.\ B {\bf 625}, 212 (2005).
%  [hep-ph/0507025].

\bibitem{Kou:2005gt}
  E.~Kou and O.~Pene,
  %``Suppressed decay into open charm for the Y(4260) being an hybrid,''
  Phys.\ Lett.\ B {\bf 631}, 164 (2005).
%  [hep-ph/0507119].

\bibitem{Close:2005iz}
  F.~E.~Close and P.~R.~Page,
  %``Gluonic charmonium resonances at BaBar and BELLE?,''
  Phys.\ Lett.\ B {\bf 628}, 215 (2005).
%  [hep-ph/0507199].


%\cite{Maiani:2014aja}
\bibitem{Maiani:2014aja}
  L.~Maiani, F.~Piccinini, A.~D.~Polosa and V.~Riquer,
  %``The Z(4430) and a New Paradigm for Spin Interactions in Tetraquarks,''
  Phys.\ Rev.\ D {\bf 89}, 114010 (2014)
  [arXiv:1405.1551 [hep-ph]].


%\cite{Faccini:2013lda}
\bibitem{Faccini:2013lda}
  L.~Maiani, V.~Riquer, R.~Faccini, F.~Piccinini, A.~Pilloni and A.~D.~Polosa,
  %``A $J^{PG}=1^{++}$ Charged Resonance in the $Y(4260) \to \pi^+ \pi^- J/\psi$ Decay?,''
  Phys.\ Rev.\ D {\bf 87}, no. 11, 111102 (2013)
  [arXiv:1303.6857 [hep-ph]].


%\cite{Brambilla:2010cs}
\bibitem{Brambilla:2010cs}
  N.~Brambilla, S.~Eidelman, B.~K.~Heltsley, R.~Vogt, G.~T.~Bodwin, E.~Eichten, A.~D.~Frawley and A.~B.~Meyer {\it et al.},
  %``Heavy quarkonium: progress, puzzles, and opportunities,''
  Eur.\ Phys.\ J.\ C {\bf 71}, 1534 (2011)  [arXiv:1010.5827 [hep-ph]].


\bibitem{italians}
R.~Faccini, A.~Pilloni and A.~D.~Polosa,
  %``Exotic Heavy Quarkonium Spectroscopy: A Mini-review,''
  Mod.\ Phys.\ Lett.\ A {\bf 27} (2012) 1230025


\bibitem{Tornqvist:2004qy}
  N.~A.~T\"ornqvist,
  %``Isospin breaking of the narrow charmonium state of Belle at 3872-MeV as a deuson,''
  Phys.\ Lett.\ B {\bf 590}, 209 (2004)
  [hep-ph/0402237].


\bibitem{Fleming:2007rp}
  S.~Fleming, M.~Kusunoki, T.~Mehen and U.~van Kolck,
  %``Pion interactions in the $X(3872)$,''
  Phys.\ Rev.\ D {\bf 76}, 034006 (2007)
  [hep-ph/0703168].

%\cite{Thomas:2008ja}
\bibitem{Thomas:2008ja}
  C.~E.~Thomas and F.~E.~Close,
  %``Is X(3872) a molecule?,''
  Phys.\ Rev.\ D {\bf 78}, 034007 (2008)
  [arXiv:0805.3653 [hep-ph]].


\bibitem{Ding:2008gr}
  G.-J.~Ding,
  %``Are Y(4260) and Z+(2) are D(1) D or D(0) D* Hadronic Molecules?,''
  Phys.\ Rev.\ D {\bf 79}, 014001 (2009).
%  [arXiv:0809.4818 [hep-ph]].

%\cite{Lee:2009hy}
\bibitem{Lee:2009hy}
  I.~W.~Lee, A.~Faessler, T.~Gutsche and V.~E.~Lyubovitskij,
  %``X(3872) as a molecular DD* state in a potential model,''
  Phys.\ Rev.\ D {\bf 80}, 094005 (2009)
  [arXiv:0910.1009 [hep-ph]].

%\cite{Dong:2009yp}
\bibitem{Dong:2009yp}
  Y.~Dong, A.~Faessler, T.~Gutsche, S.~Kovalenko and V.~E.~Lyubovitskij,
  %``X(3872) as a hadronic molecule and its decays to charmonium states and pions,''
  Phys.\ Rev.\ D {\bf 79}, 094013 (2009)
  [arXiv:0903.5416 [hep-ph]].



%\cite{Stapleton:2009ey}
\bibitem{Stapleton:2009ey}
  E.~Braaten and J.~Stapleton,
  %``Analysis of J/psi pi+ pi- and D0 anti-D0 pi0 Decays of the X(3872),''
  Phys.\ Rev.\ D {\bf 81}, 014019 (2010)
  [arXiv:0907.3167 [hep-ph]].

%\cite{Gamermann:2009fv}
\bibitem{Gamermann:2009fv}
  D.~Gamermann and E.~Oset,
  %``Isospin breaking effects in the X(3872) resonance,''
  Phys.\ Rev.\ D {\bf 80}, 014003 (2009)
  [arXiv:0905.0402 [hep-ph]].


\bibitem{Wang:2013cya}
  Q.~Wang, C.~Hanhart and Q.~Zhao,
  %``Decoding the riddle of Y(4260) and $Z_c(3900)$,''
  Phys.\ Rev.\ Lett.\  {\bf 111}, 132003 (2013).
%  [arXiv:1303.6355 [hep-ph]].

  %\cite{Wang:2013kra}
\bibitem{Wang:2013kra}
  Q.~Wang, M.~Cleven, F.-K.~Guo, C.~Hanhart, U.-G.~Mei{\ss}ner, X.~-G.~Wu and Q.~Zhao,
  %``Y(4260): hadronic molecule versus hadro-charmonium interpretation,''
  Phys.\ Rev.\ D {\bf 89}, 034001 (2014)
  [arXiv:1309.4303 [hep-ph]].
  %%CITATION = ARXIV:1309.4303;%%
  %2 citations counted in INSPIRE as of 11 Feb 2014



%\cite{Mehen:2011ds}
\bibitem{Mehen:2011ds}
  T.~Mehen and R.~Springer,
  %``Radiative Decays $X(3872) \to \psi(2S)\gamma$ and $\psi(4040) -> X(3872)\gamma$ in Effective Field Theory,''
  Phys.\ Rev.\ D {\bf 83}, 094009 (2011)  [arXiv:1101.5175 [hep-ph]].

\bibitem{Bondar:2011ev}
  A.~E.~Bondar, A.~Garmash, A.~I.~Milstein, R.~Mizuk and M.~B.~Voloshin,
  %``Heavy quark spin structure in $Z_b$ resonances,''
  Phys.\ Rev.\ D {\bf 84}, 054010 (2011)
  [arXiv:1105.4473 [hep-ph]].

\bibitem{Nieves:2011vw}
  J.~Nieves and M.~P.~Valderrama,
  %``Deriving the existence of $B\bar{B}^*$ bound states from the X(3872) and Heavy Quark Symmetry,''
  Phys.\ Rev.\ D {\bf 84}, 056015 (2011)
  [arXiv:1106.0600 [hep-ph]].

\bibitem{Nieves:2012tt}
  J.~Nieves and M.~P.~Valderrama,
  %``The Heavy Quark Spin Symmetry Partners of the X(3872),''
  Phys.\ Rev.\ D {\bf 86}, 056004 (2012)
  [arXiv:1204.2790 [hep-ph]].

\bibitem{Guo:2013sya}
  F.-K.~Guo, C.~Hidalgo-Duque, J.~Nieves and M.~P.~Valderrama,
  %``Consequences of Heavy Quark Symmetries for Hadronic Molecules,''
  Phys.\ Rev.\ D {\bf 88}, 054007 (2013)
  [arXiv:1303.6608 [hep-ph]].

\bibitem{Guo:2013zbw}
  F.-K.~Guo, C.~Hanhart, U.-G.~Mei{\ss}ner, Q.~Wang and Q.~Zhao,
  %``Production of the X(3872) in charmonia radiative decays,''
  Phys.\  Lett.\  B {\bf 725}, 127 (2013)
  [arXiv:1306.3096 [hep-ph]].

\bibitem{Mehen:2013mva}
  T.~Mehen and J.~Powell,
  %``Line shapes in  with Z(10610) and Z(10650) using effective field theory,''
  Phys.\ Rev.\ D {\bf 88},  034017 (2013)
  [arXiv:1306.5459 [hep-ph]].

\bibitem{He:2013nwa}
  J.~He, X.~Liu, Z.~F.~Sun and S.~L.~Zhu,
  %``$Z_c(4025)$ as the hadronic molecule with hidden charm,''
  Eur.\ Phys.\ J.\ C {\bf 73}, 2635 (2013)
  [arXiv:1308.2999 [hep-ph]].

\bibitem{Liu:2014eka}
  X.~H.~Liu, L.~Ma, L.~P.~Sun, X.~Liu and S.~L.~Zhu,
  %``Resolving the puzzling decay patterns of charged $Z_c$ and $Z_b$ states,''
  Phys.\ Rev.\ D {\bf 90}, 074020 (2014)
  [arXiv:1407.3684 [hep-ph]].

%\cite{Bugg:2004rk}
\bibitem{Bugg:2004rk}
  D.~V.~Bugg,
  %``Reinterpreting several narrow `resonances' as threshold cusps,''
  Phys.\ Lett.\ B {\bf 598}, 8 (2004)
  [hep-ph/0406293].


%\cite{Bugg:2011jr}
\bibitem{Bugg:2011jr}
  D.~V.~Bugg,
  %``An Explanation of Belle states $Z_b(10610) and Z_b(10650),''
  Europhys.\ Lett.\  {\bf 96}, 11002 (2011)  [arXiv:1105.5492 [hep-ph]].

%\cite{Chen:2011pv}
\bibitem{Chen:2011pv}
  D.~Y.~Chen and X.~Liu,
  %``$Z_b(10610)$ and $Z_b(10650)$ structures produced by the initial single pion emission in the $\Upsilon(5S)$ decays,''
  Phys.\ Rev.\ D {\bf 84}, 094003 (2011)
  [arXiv:1106.3798 [hep-ph]].


%\cite{Chen:2011xk}
\bibitem{Chen:2011xk}
  D.~Y.~Chen and X.~Liu,
  %``Predicted charged charmonium-like structures in the hidden-charm dipion decay of higher charmonia,''
  Phys.\ Rev.\ D {\bf 84}, 034032 (2011)
  [arXiv:1106.5290 [hep-ph]].

%\cite{Chen:2011pu}
\bibitem{Chen:2011pu}
  D.~Y.~Chen, X.~Liu and T.~Matsuki,
  %``Charged bottomonium-like structures in the hidden-bottom dipion decays of $\Upsilon(11020)$,''
  Phys.\ Rev.\ D {\bf 84}, 074032 (2011)
  [arXiv:1108.4458 [hep-ph]].

%\cite{Chen:2013coa}
\bibitem{Chen:2013coa}
  D.~Y.~Chen, X.~Liu and T.~Matsuki,
  %``Reproducing the $Z_c(3900)$ structure through the initial-single-pion-emission mechanism,''
  Phys.\ Rev.\ D {\bf 88}, 036008 (2013)
  [arXiv:1304.5845 [hep-ph]].

%\cite{Chen:2013wca}
\bibitem{Chen:2013wca}
  D.~Y.~Chen, X.~Liu and T.~Matsuki,
  %``Predictions of Charged Charmoniumlike Structures with Hidden-Charm and Open-Strange Channels,''
  Phys.\ Rev.\ Lett.\  {\bf 110}, 232001 (2013)
  [arXiv:1303.6842 [hep-ph]].

%\cite{Swanson:2014tra}
\bibitem{Swanson:2014tra}
  E.~S.~Swanson,
  %``$Z_b$ and $Z_c$ Exotic States as Coupled Channel Cusps,''
  arXiv:1409.3291 [hep-ph].


\bibitem{Batley:2000zz}
  J.~R.~Batley, A.~J.~Culling, G.~Kalmus, C.~Lazzeroni, D.~J.~Munday, M.~W.~Slater, S.~A.~Wotton and R.~Arcidiacono {\it et al.},
  %``Determination of the S-wave pi pi scattering lengths from a study of K+- ---> pi+- pi0 pi0 decays,''
  Eur.\ Phys.\ J.\ C {\bf 64}, 589 (2009)
  [arXiv:0912.2165 [hep-ex]].

\bibitem{Gasser:2011ju}
  J.~Gasser, B.~Kubis and A.~Rusetsky,
  %``Cusps in K --> 3pi decays: a theoretical framework,''
  Nucl.\ Phys.\ B {\bf 850}, 96 (2011)
  [arXiv:1103.4273 [hep-ph]].

\bibitem{Wu:2011yx}
  J.~J.~Wu, X.~H.~Liu, Q.~Zhao and B.~S.~Zou,
  %``The Puzzle of anomalously large isospin violations in $\eta(1405/1475)\to
  %3\pi$,''
  Phys.\ Rev.\ Lett.\  {\bf 108}, 081803 (2012)
  [arXiv:1108.3772 [hep-ph]].

%\cite{Wu:2012pg}
\bibitem{Wu:2012pg}
  X.~G.~Wu, J.~J.~Wu, Q.~Zhao and B.~S.~Zou,
  %``Understanding the property of $\eta(1405/1475)$ in the $J/\psi$ radiative decay,''
  Phys.\ Rev.\ D {\bf 87}, 014023 (2013)
  [arXiv:1211.2148 [hep-ph]].
 

\bibitem{Szczepaniak:2015eza}
A.~P.~Szczepaniak,
  %``Triangle Singularities and XYZ Quarkonium Peaks,''
  arXiv:1501.01691 [hep-ph].


\bibitem{Wang:2013hga}
Q.~Wang, C.~Hanhart and Q.~Zhao,
  %``Systematic study of the singularity mechanism in heavy quarkonium decays,''
  Phys.\ Lett.\ B {\bf 725}, 106 (2013)
  [arXiv:1305.1997 [hep-ph]].


\bibitem{Schmid:1967}
  C.~Schmid,
  Phys.\ Rev.\ {\bf 154}, 1363 (1967).


\bibitem{Anisovich:1995ab}
  A.~V.~Anisovich and V.~V.~Anisovich,
  %``Rescattering effects in three particle states and the Schmid theorem,''
  Phys.\ Lett.\ B {\bf 345}, 321 (1995).

\bibitem{Danilkin:2011sh}
  I.~V.~Danilkin, V.~D.~Orlovsky and Y.~A.~Simonov,
  %``Hadron interaction with heavy quarkonia,''
  Phys.\ Rev.\ D {\bf 85}, 034012 (2012)
  [arXiv:1106.1552 [hep-ph]].

\end{thebibliography}
\end{document}